\documentclass[a4paper]{spicawStyle}
\usepackage{graphicx,natbib}

\begin{document}

\title{{\it AKARI} Far-Infrared All Sky Survey}

\author{Y. Doi, \inst{1} \and M. Etxaluze Azkonaga\inst{2,3} \and G. White\inst{2,3} \and E. Figuered\inst{4} \and Y. Chinone\inst{5} \and M. Hattori\inst{5} \and N. Ikeda\inst{6} \and Y. Kitamura\inst{6} \and S. Komugi\inst{6} \and T. Nakagawa\inst{6} \and C. Yamauchi\inst{6} \and Y. Matsuoka\inst{7} \and H. Kaneda\inst{7} \and M. Kawada\inst{7} \and H. Shibai\inst{8} \and the AKARI team}

\institute{
Department of Earth Science and Astronomy, Graduate School of Arts and Sciences, The University of Tokyo, Komaba 3-8-1, Meguro, Tokyo, 153-8902, Japan
\and
Department of Physics \& Astronomy, The Open University, Milton Keynes, MK7 6BJ, United Kingdom
\and
Space Science and Technology Dept., The Rutherford Appleton Laboratory, Didcot
OX11 0QX, United Kingdom
\and
Instituto de Astronomiac, Geof\'isica e Ci\^encias Atmosf\'ericas, Departamento de Astronomica, Universidade de S\~ao Paulo, Rua do Mat\~ao 1226, Cidade Universitaria 05508-900, S\~ao Paulo SP, Brazil
\and
Astronomical Institute, Graduate School of Science, Tohoku university, Aramaki, Aoba-ku, Sendai, 980-8578, Japan
\and
Institute of Space and Astronautical Science, Japan Aerospace Exploration Agency, Yoshinodai 3-1-1, Sagamihara, Kanagawa, 229-8510, Japan
\and
Department of Physics and Astrophysics, Nagoya University, Furocho, Chikusa, Nagoya, 464-8602, Japan
\and
Department of Earth and Space Science, Graduate School of Science, Osaka University, 1-1 Machikaneyama, Toyonaka 560-0043, Osaka, Japan}

\maketitle 

\begin{abstract}

We demonstrate the capability of {\it AKARI} for mapping diffuse far-infrared emission and achieved reliability of all-sky diffuse map.
We have conducted an all-sky survey for more than 94 \% of the whole sky during cold phase of {\it AKARI} observation in 2006 Feb. -- 2007 Aug.
The survey in far-infrared waveband covers 50 $\mu$m -- 180 $\mu$m with four bands centered at 65 $\mu$m, 90 $\mu$m, 140 $\mu$m, and 160 $\mu$m
and spatial resolution of 30\arcsec\ -- 40\arcsec (FWHM).
This survey has allowed us to make a revolutionary improvement compared to the IRAS survey that was conducted in 1983 in both spatial resolution and sensitivity after more than a quarter of a century.
Additionally, it will provide us the first all-sky survey data with high-spatial resolution beyond 100 $\mu$m.
Considering its extreme importance of the {\it AKARI} far-infrared diffuse emission map, we are now investigating carefully the quality of the data for possible release of the archival data.
Critical subjects in making image of diffuse emission from detected signal are the transient response and long-term stability of the far-infrared detectors.
Quantitative evaluation of these characteristics is the key to achieve sensitivity comparable to or better than that for point sources ($<20$ -- 95 [MJy/sr]).
We describe current activities and progress that are focused on making high quality all-sky survey images of the diffuse far-infrared emission.

\keywords{Surveys -- Atlases -- ISM: general -- Galaxy: general -- Galaxies: general -- Infrared: ISM -- Infrared: galaxies}
\end{abstract}

\section{{\it AKARI} satellite and the {\it AKARI} all-sky survey}
\label{doiy3_sec:intro}

{\it AKARI} is the first Japanese satellite dedicated for infrared (IR) astronomy \citep{doiy3:murakami07}.
The satellite was launched in February 2006.
Its telescope with a primary mirror of a diameter of $\phi = 68.5$ cm as well as the focal plane instruments were cooled down with liquid Helium cryogen and mechanical JT coolers as a support \citep{doiy3:nakagawa07}.
During its cold operational phase with liquid Helium (cold phase; March 2006 -- August 2007), infrared observations in 2 -- 180 $\mu$m wavelength were carried out.

One of the key observations of the {\it AKARI} mission was to conduct all-sky survey in mid- and far-IR (MIR and FIR) wavelengths centered at 9 $\mu$m, 18 $\mu$m, 65 $\mu$m, 90 $\mu$m, 140 $\mu$m, and 160 $\mu$m.
In this contribution, we will discuss FIR part of the survey observation.

\begin{figure}[bt]
  \begin{center}
    \includegraphics[width=7 cm]{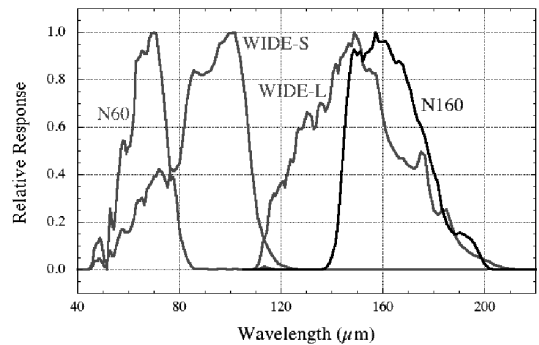}
  \end{center}
  \caption{Spectral responses of the {\it AKARI} FIR four bands centered at 65 $\mu$m (N60), 90 $\mu$m (WIDE-S), 140 $\mu$m (WIDE-L), and 160 $\mu$m (N160). Note that {\it AKARI} has continuous wavebands covering 50--180 $\mu$m so that we can make precise evaluation of total FIR intensity from inband flux of {\it AKARI} observations (see $\S$\ref{doiy3_sec:discussion}).}
\label{doiy3_fig:spectra}
\end{figure}

Specifications of the FIR wavebands are shown in Table~\ref{doiy3_tab:spec}.
\begin{table*}[bht]
  \caption{Specifications of {\it AKARI} FIR detectors (\citeauthor{doiy3:kawada07} \citeyear{doiy3:kawada07}, \citeauthor{doiy3:shirahata09} \citeyear{doiy3:shirahata09}).}
  \label{doiy3_tab:spec}
  \begin{center}
    \leavevmode
    \footnotesize
    \begin{tabular}[h]{lccccl}
      \hline \\[-5pt]
      Band name & N60 & WIDE-S & WIDE-L & N160\\[+5pt]
      \hline \\[-5pt]
      Center wavelength & 65 & 90 & 140 & 160 & [$\mu$m]\\
      Wavelength range & 50 -- 80 & 60 -- 110 & 110 -- 180 & 140 -- 180 & [$\mu$m]\\
      Array format & $20\times 2$ & $20\times 3$ & $15\times 3$ & $15\times 2$ & [pixels]\\
      Pixel scale\footnotemark[1] & \multicolumn{2}{c}{$26\arcsec \negthinspace\negthinspace .8 \times 26\arcsec \negthinspace\negthinspace .8$} & \multicolumn{2}{c}{$44\arcsec \negthinspace\negthinspace .2 \times 44\arcsec \negthinspace\negthinspace .2$} & \\
      Pixel pitch\footnotemark[1] & \multicolumn{2}{c}{$29\arcsec \negthinspace\negthinspace .5 \times 29\arcsec \negthinspace\negthinspace .5$} & \multicolumn{2}{c}{$49\arcsec \negthinspace\negthinspace .1 \times 49\arcsec \negthinspace\negthinspace .1$} & \\
      FWHM & $32\arcsec \negthinspace\negthinspace .05 \pm 0\arcsec \negthinspace\negthinspace .10$ & $30\arcsec \negthinspace\negthinspace .17 \pm 0\arcsec \negthinspace\negthinspace .08$ & $40\arcsec \negthinspace\negthinspace .85 \pm 0\arcsec \negthinspace\negthinspace .10$ & $38\arcsec \negthinspace\negthinspace .23 \pm 0\arcsec \negthinspace\negthinspace .15$ & \\
      Detector device & \multicolumn{2}{c}{Monolithic Ge:Ga array} & \multicolumn{2}{c}{Stressed Ge:Ga array} & \\
      \hline \\[-8pt]
      \multicolumn{6}{l}{\footnotesize{\footnotemark[1]At the array center.}}\\
      \end{tabular}
  \end{center}
\end{table*}
Four wavebands continuously cover 50 -- 180 $\mu$m as shown in Figure~\ref{doiy3_fig:spectra} (also see \citeauthor{doiy3:kawada07} \citeyear{doiy3:kawada07}).
During the survey observation, the telescope direction is being kept opposite to the earth centre.
Since the {\it AKARI} satellite was launched into a sun-synchronous polar orbit and is revolving around the earth in every 100 minutes, strips of the sky along a great circle with 8\arcmin -- 12\arcmin\ width, which is the width of detector arrays, are scanned with a constant scan speed of 3\arcmin\negthinspace.6/sec.

Due to the yearly revolution of the earth, the scan direction is shifted 4\arcmin~per satellite revolution in longitudinal direction or in the cross-scan direction.
This shift angle corresponds to about half or one-third of the detector widths so that we can observe a region in the sky twice or more with a continuous survey observation.

The whole sky is covered with half-a-year continuous observation except for the regions that cannot be surveyed due to moon-shine, heavy bombardment of high-energy particles due to anomalous geomagnetic field above South America (South-Atlantic Anomaly: SAA), and so on.

Pointed observations of specific sources and/or regions also hinder the survey observation by blocking 30-minute observational time per each observation.
In these observations, FIR 4 detectors map the sky with a scan speed of 7\arcsec\negthinspace.5/sec, 15\arcsec/sec, or 30\arcsec/sec.
These speeds are much slower (slow-scan observations) than that of the survey observation, so that the data have better quality than the survey data with much more spatial samplings and smaller artifacts due to non-linear behaviour of the detectors (slow-response; see also $\S$\ref{doiy3_sec:analysis}).
We make comparisons between survey data and slow-scan data taken at the same region of the sky to evaluate the reliability of the data in $\S$\ref{doiy3_sec:cal}.

\begin{figure}[ht]
  \begin{center}
    \includegraphics[width=7 cm]{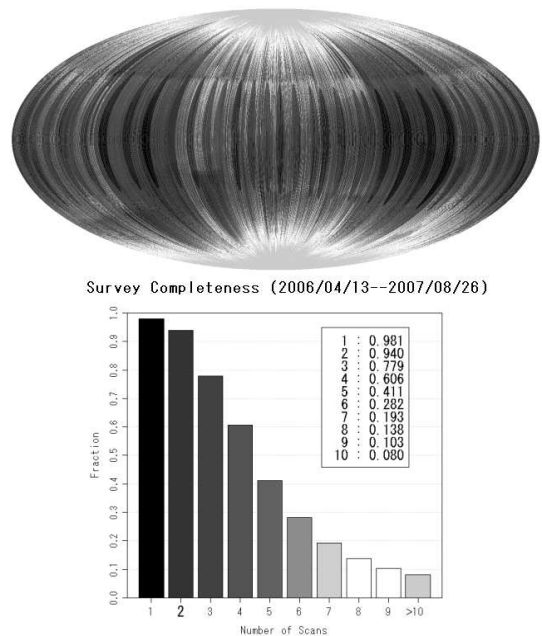}
  \end{center}
  \caption{Sky coverage of the {\it AKARI} all-sky survey in ecliptic coordinates (top panel). Cumulative fraction of the sky coverage as a function of the scan number is shown in the bottom panel.}
\label{doiy3_fig:cov}
\end{figure}

Observational gaps in the first six-month operation were compensated in later periods within the limit of satellite attitude control; cross-scan offset $< \pm 1\degr$.
After about 17 months of survey period (cold phase), more than 94 \% of the sky have been observed twice or more so that we have conducted a FIR all-sky survey that virtually covers the whole sky
as shown in Fugure~\ref{doiy3_fig:cov}.

\section{data analysis}
\label{doiy3_sec:analysis}

We are now developing an image processing program.
Here we describe the current status of the development.

Direction of detector FOV is determined by post-flight analysis of NIR star camera.
Long-term drift of the detector response is corrected by calibration sequence using a cold shutter (dark and controlled photon influx by internal calibration light) as well as periodic flashing of the calibration light in every one-minute during the observation (\citeauthor{doiy3:kawada07} \citeyear{doiy3:kawada07}, \citeauthor{doiy3:yamamura09} \citeyear{doiy3:yamamura09}).

In addition to the long-term drift, non-linear behaviour of the detector is another major cause to distort the detector time-line signal \citep{doiy3:kaneda09}.
The distortion is corrected by applying time-line numerical filter \citep{doiy3:doi09}.

The resultant data are used to process an image.
A preliminary image is created with a rather large grid size as compared to the PSF so that the image data have sufficiently large samplings in each bin ensuring a reliable removal of outliers and/or smoothing out correction residuals of the detector sensitivity drift.
By referring to this preliminary image, anomalous data, including spikes due to cosmic ray hitting and so on, are removed from the raw data.
Additional correction on the long-term drift of the detector responsivity is made by adding a constant offset to each time-line data strip by referring the preliminary image.
This correction works as a destriping process.
A final image is then processed using the resultant data.

\begin{figure}[bt]
  \begin{center}
    \includegraphics[width=7 cm]{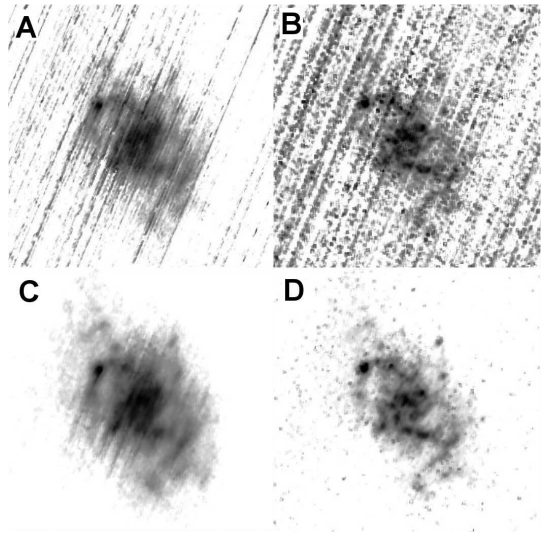}
  \end{center}
  \caption{FIR 140 $\mu$m survey images of M33 for illustrating the efficiency of the slow-response correction and the destriping; A: original image, B: with slow-response correction, C: with destriping, D: the resultant image with the two corrections. Image sizes are $1\degr \times 1\degr$.}
\label{doiy3_fig:M33}
\end{figure}

Figure~\ref{doiy3_fig:M33} is a set of example images of M33.
Effects of both corrections of the non-linear detector behaviour and the long-term responsivity drift are shown.
Judging from the figure, it can be concluded that both the corrections significantly improve the quality of the final product image.

\section{Calibration and reliability of the image data}
\label{doiy3_sec:cal}

Here we describe our current status of data calibration and achieved reliability of image data.
To evaluate the reliability, first we compare the data with slow-scan data (see $\S$\ref{doiy3_sec:intro}) taken at the same region of the sky.

\begin{figure}[bt]
  \begin{center}
    \includegraphics[width=9 cm]{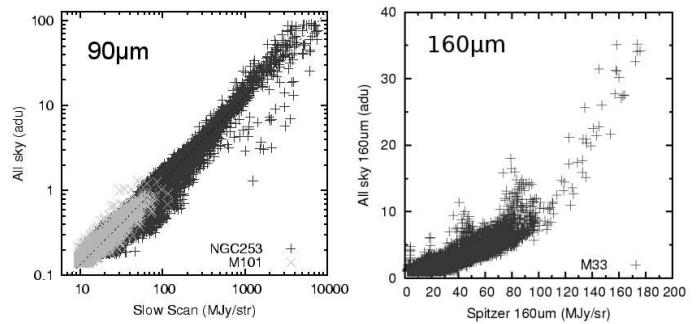}
  \end{center}
  \caption{Cross-correlations between {\it AKARI} survey data with other data. The left panel shows a comparison of 90$\mu$m survey and slow-scan data of NGC 253 and M101. The right panel shows a comparison of 160$\mu$m M33 survey data with the data taken by Spitzer.}
\label{doiy3_fig:N253}
\end{figure}

Figure~\ref{doiy3_fig:N253} shows comparisons of NGC 253 and M101 data.
Reasonably good correlations are found between the survey and the slow-scan data, showing that the slow-response correction and destriping work well for improving the quality of the survey data.
The relative scatter of the correlation is estimated to be $\sim 50\%$, which can be regarded as the relative accuracy of the survey data.

A comparison of the survey data with Spitzer observation of M33 is also shown in Figure~\ref{doiy3_fig:N253}.
The relative scatter of the correlation is estimated as $\sim 50\%$, which is consistent with the aforementioned relative scatter of the correlation between the survey and the slow-scan data.
Since we need to evaluate the relative/absolute accuracy for other regions and assess the stability of the calibration, which might degrade the estimated accuracy to some extent, we conclude that the currently achieved absolute and relative accuracy of the {\it AKARI} survey data should be better than $\pm 100\%$.
Further investigation is required especially to improve the correction of the long-term sensitivity drift.

\section{Discussion and Summary}
\label{doiy3_sec:discussion}

\begin{figure}[ht]
  \begin{center}
    \includegraphics[width=7 cm]{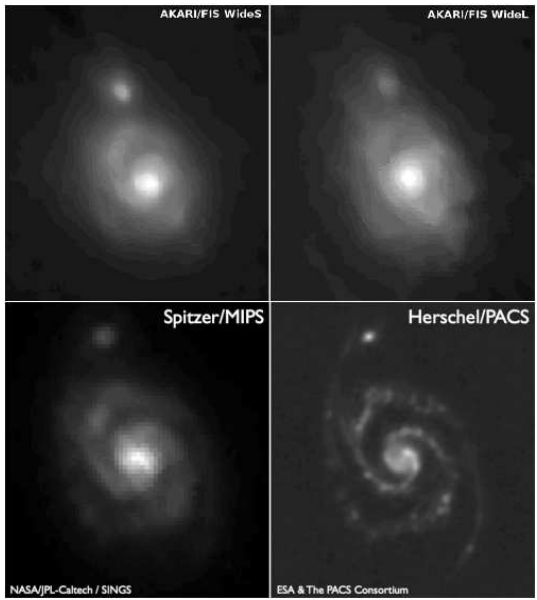}
  \end{center}
  \caption{M51 images observed with {\it AKARI}, Spitzer, and Herschel; top-left: {\it AKARI} 90 $\mu$m survey image, top-right: {\it AKARI} 140 $\mu$m survey image; bottom images are taken from ESA Herschel press release on 19 June 2009; bottom-left: Spitzer/MIPS 160 $\mu$m image (NASA/JPL-Caltech/SINGS), bottom-right: Herschel/PACS 160 $\mu$m image (ESA \& the PACS consortium).}
\label{doiy3_fig:M51}
\end{figure}

One of the key characteristics of the {\it AKARI} FIR all-sky survey is its good spatial resolution of 30\arcsec\ -- 40\arcsec\ (Table~\ref{doiy3_tab:spec}) for covering the whole sky.
This advantage is illustrated in Figure \ref{doiy3_fig:M51} as the spatial resolution of {\it AKARI} is nearly comparable to that of Spitzer.

\begin{figure}[ht]
  \begin{center}
    \includegraphics[width=7 cm]{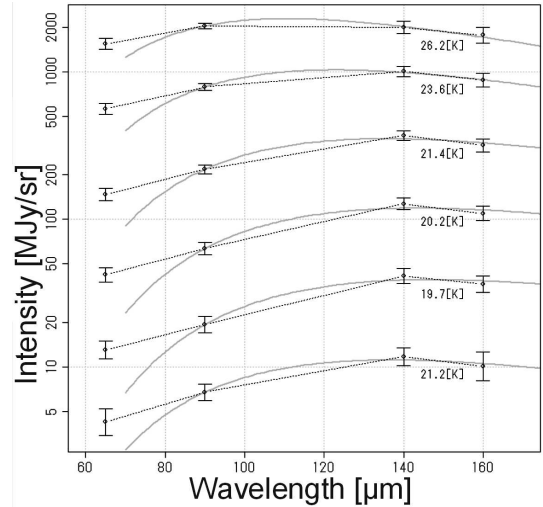}
  \end{center}
  \caption{FIR spectra observed in LMC. Average spectra in intensity slices are shown. Gray lines are gray-body single temperature fittings with a spectral index $\beta = 1$.}
\label{doiy3_fig:LMC}
\end{figure}

Wide and continuous spectral coverage of 50 -- 180 $\mu$m (Table~\ref{doiy3_tab:spec}, Figure~\ref{doiy3_fig:spectra}) is another important advantage of the {\it AKARI} FIR all-sky survey.
Observed spectra, which are averaged in individual intensity bins, are shown in Figure~\ref{doiy3_fig:LMC}.
Single-temperature gray-body spectra are also shown.
Four spectral bands cover both sides of the spectral peak in shorter wavelengths (65 $\mu$m and 90 $\mu$m) and longer wavelengths (140 $\mu$m and 160 $\mu$m) enabling us to make precise evaluation of colour temperature, as shown in the figure.

It is important to notice that excess emission is prominent in the 65 $\mu$m band for all the intensity bins, showing significant contribution of emission from very small grains (VSGs; \citeauthor{doiy3:dbp90} \citeyear{doiy3:dbp90}).
Thus 90 $\mu$m / 140 $\mu$m intensity ratio is a good indicator of the colour temperature of big grains (BGs), which are in a thermal equilibrium with interstellar radiation field (\citeauthor{doiy3:dbp90} \citeyear{doiy3:dbp90}), although possible excess emission even in 90 $\mu$m band is found at the lowest intensity bin as the estimated colour temperature shows a small rise comparing to higher intensity bins.

Since its continuous coverage of the {\it AKARI} bands in 50 -- 180 $\mu$m with 90 $\mu$m and 140 $\mu$m bands (WideS band and WideL band; Figure~\ref{doiy3_fig:spectra}), total intensity in 50 -- 180 $\mu$m ($I_{50-180}$) can be estimated simply by summing up 90 $\mu$m intensity $I_\nu ({\rm WideS})$ and 140 $\mu$m intensity $I_\nu ({\rm WideL})$ as follows:
\[
I_{50-180} = \Delta \nu I_\nu ({\rm WideS}) + \Delta \nu I_\nu ({\rm WideL}).
\]

\citet{doiy3:doi09} make numerical evaluation of the fraction between $I_{50-180}$ and $I_{30-800}$.
The fraction is estimated to be 0.4 -- 0.7 in variety of conditions of interstellar space, leading them to conclude that the {\it AKARI} total FIR intensity, $I_{50-180}$, is a good indicator of the total infrared intensity (TIR).

Takeuchi et al. (\citeyear{doiy3:takeuchi09}) confirmed this indication by finding good luminosity correlation between $L_{50-180}$ and $L_{\rm TIR}$ for nearby galaxies, expressed by the following equation: 
\[
{\rm log} L_{\rm TIR} = 0.964\ {\rm log} L^{\rm 2band}_{\rm AKARI} + 0.814,
\]\[
r = 0.989.
\]

It is widely accepted that $I_{\rm TIR}$ is a good indicator of star-formation activities (e.g. \citeauthor{doiy3:kennicutt98} \citeyear{doiy3:kennicutt98}) so that the all-sky data of the {\it AKARI} FIR survey can be a powerful tool to investigate the star-formation activities in the various sources including near-by star-formation regions as well as distant galaxies.

As a result, the {\it AKARI} FIR all-sky survey can be a new basic database that is improved from the IRAS survey with its higher spatial resolution and wider spectrum coverage.
It can be utilized as a fundamental database for variety of astronomical studies as well as a pilot survey for the current and the future infrared astronomical missions including Herschel, Planck and SPICA.

The data will be released among the {\it AKARI} project members after the completion of the calibration ($\S$\ref{doiy3_sec:cal}) and
the public will have access to the data through collaborations with project members.
We are planning to release the data to the public after some proprietary time as the {\it AKARI} FIR point source catalogue \citep{doiy3:yamamura09}.

\end{document}